\documentclass{article}
\usepackage[square,sort,comma,numbers]{natbib}
\usepackage[margin=3cm, top=2cm]{geometry}

\usepackage{graphicx} 
\usepackage{xcolor}
\definecolor{royalazure}{rgb}{0.0, 0.22, 0.86}
\definecolor{royalblue}{rgb}{0.04,0.33,0.64 }
\usepackage[
    colorlinks,
    citecolor=royalazure,
    filecolor=royalazure,
    linkcolor=royalazure,
    urlcolor=blue
]{hyperref}
\usepackage[font=small,labelfont=bf]{caption}
\usepackage{placeins}
\usepackage{authblk}

\usepackage{subcaption}
\title{Language models in molecular discovery}
\author[1]{Nikita Janakarajan}
\author[2]{Tim Erdmann}
\author[2]{Sarath Swaminathan}
\author[1]{Teodoro Laino}
\author[1*]{Jannis Born}

\affil[1]{IBM Research Europe, Zurich, Switzerland}
\affil[2]{IBM Research Almaden, San Jose, CA, United States}
\vspace{3mm}
\affil[*]{Corresponding author: jab@zurich.ibm.com}

\usepackage{xcolor}
\usepackage{enumitem}

\begin{document}
\vspace{-8cm}

\maketitle
\vspace{-5mm}
\begin{abstract}
    The success of language models, especially transformer-based architectures, has trickled into other domains giving rise to "scientific language models" that operate on small molecules, proteins or polymers.
    In chemistry, language models contribute to accelerating the molecule discovery cycle as evidenced by 
    promising recent findings 
    in early-stage drug discovery. 
    Here, we review the 
    role of
    language models in molecular discovery, underlining their strength in de novo drug design, property prediction and reaction chemistry.
    We highlight valuable open-source software assets thus lowering the entry barrier to the field of scientific language modeling.
    Last, we sketch a vision for future molecular design that combines a chatbot interface with access to computational chemistry tools.
    Our contribution serves as a valuable resource for researchers, chemists, and AI enthusiasts interested in understanding how language models can and will be used to accelerate chemical discovery.
\end{abstract}


\section{Introduction}
Despite technological advances constantly reshaping our understanding of biochemical processes, the chemical industry persistently faces escalating resource costs of up to 10 years and 3 billion dollar per new market release~\cite{wouters2020estimated}.
The intricacy of the problem is typically attested by an exorbitant attrition rate in \textit{in vitro} screenings~\cite{scannell2012diagnosing}, the sheer size of the chemical space~\cite{polishchuk2013estimation} and the frequency of serendipity~\cite{hargrave2012serendipity}.

Language models (LMs) emerged recently and demonstrated an astonishing ability to understand and generate human-like text~\cite{openai2023gpt4}. 
%
\begin{figure}[!ht]
\begin{center}
    \begin{subfigure}{0.42\textwidth}
      \centering
      \includegraphics[width=1\linewidth]{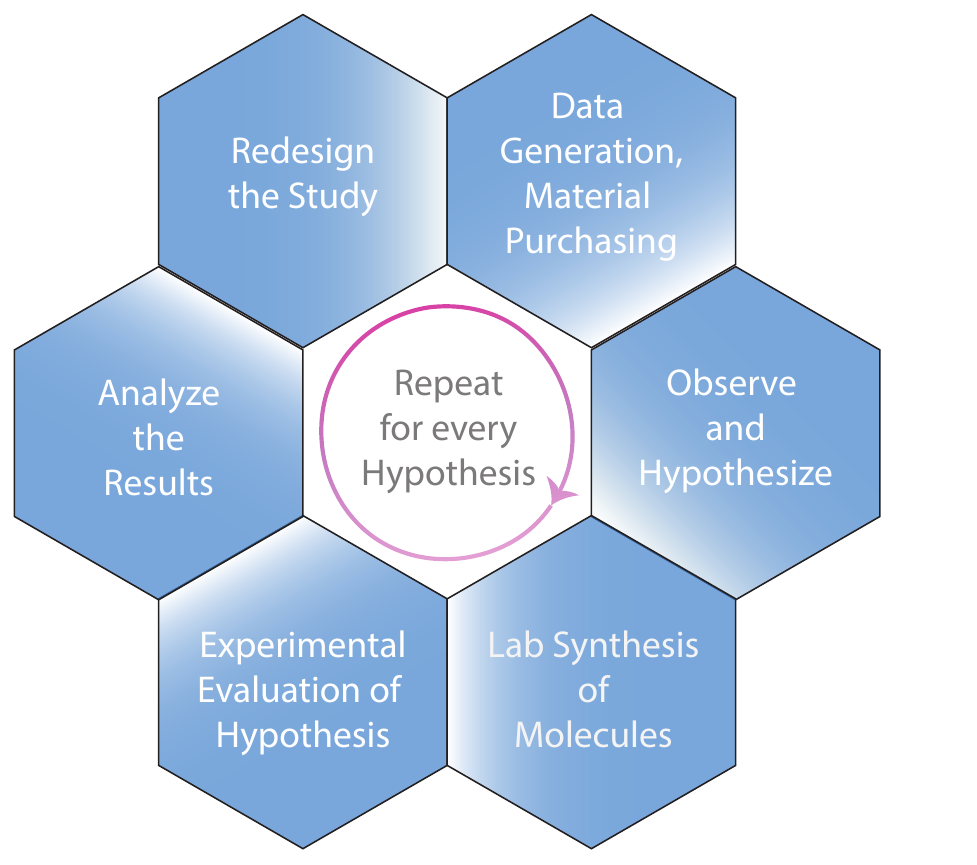}
      \caption{Classic molecular discovery.}
      \label{fig:work_sub1}
    \end{subfigure}%
    \begin{subfigure}{0.42\textwidth}
      \centering
      \includegraphics[width=1\linewidth]{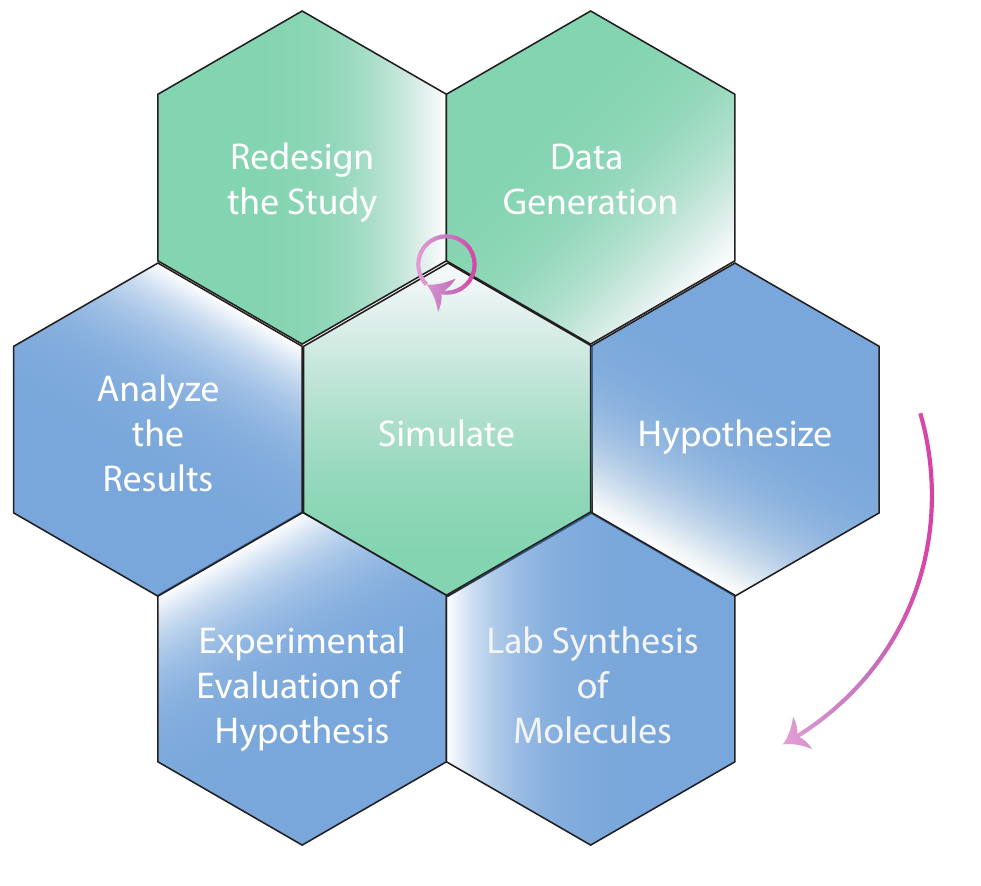}
      \caption{Accelerated molecular discovery.}
      \label{fig:work_sub2}
    \end{subfigure}
    \end{center}
    \vspace{-5mm}
    \caption{
    A comparison of molecular discovery workflows: (a) classic approach, where each hypothesis (a.k.a. molecule) requires a new experimental cycle. (b) \textit{Accelerated} molecular discovery cycle with machine-generated hypotheses and assisted validation, enabling simultaneous generation and testing of numerous molecules.
    }
    \label{fig:workflow}
\end{figure}
%
Machine learning (ML) in general and LMs in particular hold the potential to profoundly accelerate the molecular discovery cycle (see~\autoref{fig:workflow}). 
In this chapter, we explore applications of LMs to chemical design tasks.
Although LMs were originally developed for natural language, they have shown compelling results in scientific discovery settings when applied to "scientific languages", e.g., in protein folding~\cite{lin2023evolutionary} or \textit{de novo} design of small molecules~\cite{zhavoronkov2019deep}, peptides~\cite{das2021accelerated} or polymers~\cite{park2023artificial}.
But what exactly is a language model?
By definition, it is any ML model that consumes a sequence of text chunks (so-called tokens) and is capable to reason about the content of the sequence.
Since each token is essentially a vector~\cite{mikolov2013efficient}, a LM is a pseudo-discrete time series model.
Most typically, LMs learn probability distributions over sequences of words thus also facilitating the generation of new text given some input, for example in a language translation task.
While all LMs rely on neural networks, contemporary models almost exclusively leverage the Transformer architecture~\cite{vaswani2017attention}. Now, all of this begs the question -- what is the need for LMs in molecular discovery?


First, when applied to serializations of chemical entities (e.g., SMILES~\cite{weininger1988smiles}), LMs can learn highly structured representations, often even tailored for desired functional properties~\cite{gomez2018automatic}.
This allows to perform smooth and property-driven exploration of the originally deemed discrete protein or molecular space.
Another attractive feature of scientific LMs is their ability to seamlessly bridge natural and scientific languages.
This can give rise to ChatGPT-style chatbot interfaces that allow chemists to formulate their design objectives through natural language and to iteratively refine their result with an interactive agent thus potentially accomplishing complex chemical tasks more rapidly.
Here, we present an overview of the role of LMs toward accelerated molecular discovery. 
We commence with the conventional scientific discovery method and then discuss how molecular generative models can be coupled with molecular property prediction models.
Seeking for practical usability, we then present the reader with selected software tools and libraries for scientific language modeling.
We close with a vision for future molecule design that integrates natural language models into the discovery process through chatbots.

\section{Accelerated molecular discovery}

Molecule discovery, intricately linked to optimizing diverse properties in a vast space, challenges conventional scientific methods. In chemistry's Design-Make-Test-Analyze (DMTA) cycle, synthesis costs and time constraints create a bottleneck that hampers hypothesis refinement (cf.~\autoref{fig:workflow}a).
Traditional approaches are largely driven by medicinal chemists who design "molecule hypotheses" which are biased, ad-hoc and non-exhaustive. This hinders progress in addressing global issues, driving crucial necessity for an accelerated process of molecule discovery. Thus, a key challenge lies in improving speed and quality of evaluating such "molecule hypotheses" grounded on laboratory work.

Deep generative models have recently emerged as a promising tool to expedite the hypothesis/design phase in molecular discovery. However, even the most advanced molecular generative models require an efficient method for large-scale virtual screening to test their hypotheses. The \textit{accelerated molecular discovery} cycle adds a validation loop to DMTA, rapidly evaluating numerous hypotheses inexpensively (cf.~\autoref{fig:workflow}b).
This loop enhances the design-phase generative model, ensuring only promising hypotheses advance to the synthesis and physical experimentation stages.


\subsection{Molecule Representation}

\begin{figure}
    \centering
    \includegraphics[width=1\linewidth]{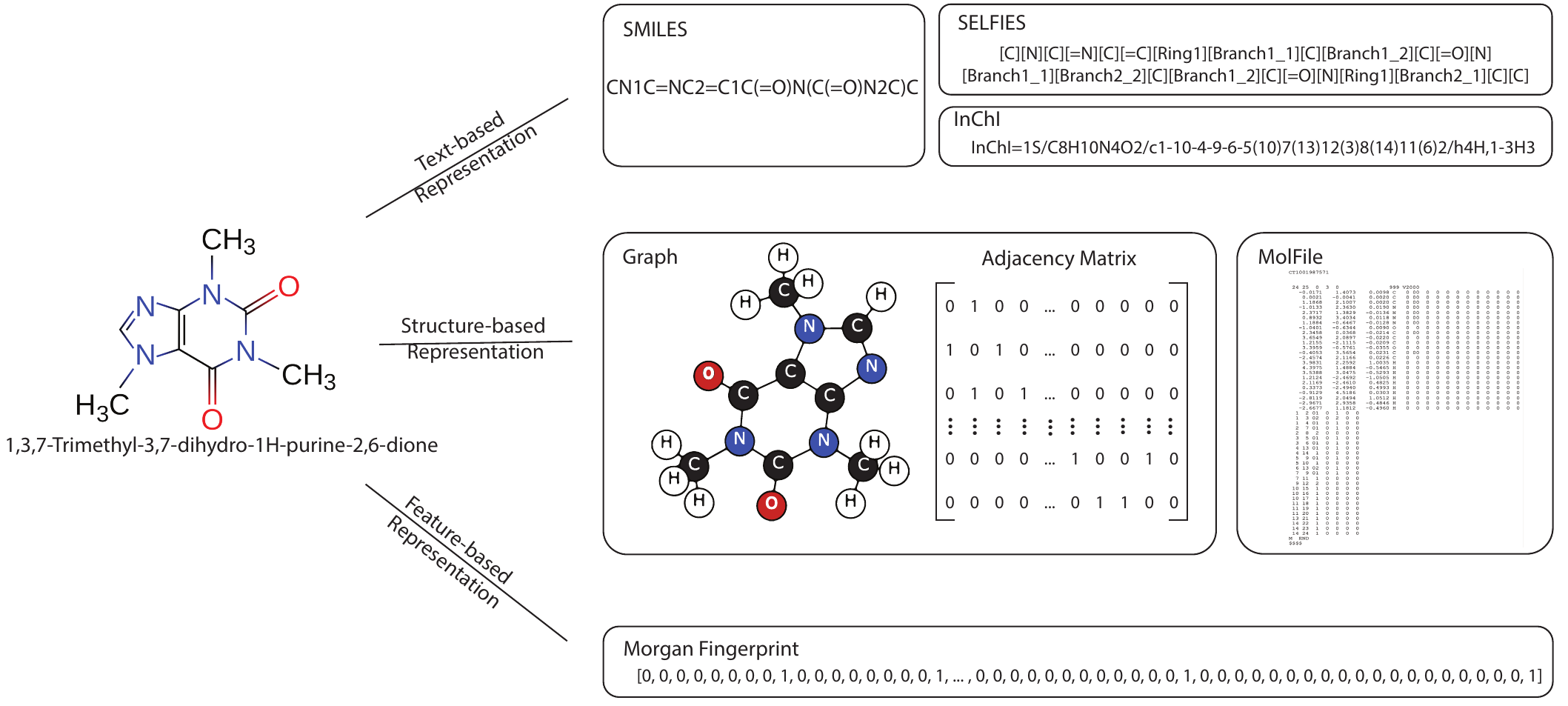}
    \caption{An illustration of popular ways of representing a chemical molecule as input to a ML model. The representations may be (a) String-based, such as SMILES, SELFIES, or InChI which use characters to represent different aspects of a molecule, (b) Structure-based, such as Graphs or MolFiles that encode connectivity and atomic position, and (c) Feature-based, such as Morgan Fingerprints, which encode local substructures as bits.}
    \label{fig:mol_rep}
\end{figure}
Data representation is critical as it determines which information is available for the model. 
As illustrated in~\autoref{fig:mol_rep}, various molecular representations exist.
Due to popularity of chemical language models (CLMs), this section focuses on text-representations of molecules.
A more focused discussion on CLMs was published by~\citet{grisoni2023chemical}.
 
\paragraph{Simplified Molecular Input Line-Entry System (SMILES)}
SMILES~\cite{weininger1988smiles} 
is a string representation made up of specific characters for atoms, bonds, branches, aromaticity, rings and stereochemistry in molecular structures. 
The character-level representation enables easy tokenization, making SMILES an ideal input for LMs.
SMILES are non-unique, so each molecule can be written as multiple SMILES strings. 
Hence, SMILES are either canonicalized or, alternatively, their multiplicity is used as data augmentation strategy~\cite{bjerrum2017smiles} which has shown performance improvement in molecular property prediction~\cite{bjerrum2017smiles,tetko2019augmentation,li2020inductive} and molecular generation~\cite{arus2019randomized,van2020gen}.
In generative modeling, a common issue is the invalidity of SMILES strings due to an uneven number of ring opening/closure symbols or bond valence violations.
SMILES strings can undergo further processing, such as kekulization or stereoinformation removal but employing canonicalized SMILES remains the most prevalent approach. 

\textbf{Tokenization} is the process of splitting a string into vectorizable units. These units are typically a single character, n-gram characters or words. 
Instead of splitting at the character level, SMILES are typically
tokenized at the atom level with regular expressions~\cite{schwaller2018found} or by additionally including positional and connectivity information, thereby acknowledging that the same atom can have different encodings based on its location in the molecular structure~\cite{ucak2023improving}. 
SMILES may also be tokenized at the substructure level, as demonstrated by SMILES Pair Encoding (SMILES-PE)~\cite{li2021smiles}. This method, inspired by byte-pair encoding, iteratively counts and merges frequently occurring SMILES token pairs until a given condition is met. Tokenization enables the creation of a vocabulary for SMILES representations.

\textbf{Vocabularies} are dictionaries mapping tokens to vectors thus serving as gateway to LMs. For LMs to learn from SMILES, tokens are vectorized, either via one-hot encodings (where each row in the binary matrix corresponds to a SMILES position and each column signifies a token). However, this discrete method results in sparse, large matrices and thus, an alluring alternative is to learn a continuous embedding for each token during training. 
This facilitates the learning of semantic relationships between tokens and enhances performance.  
Since learning good embeddings requires a lot of data, models pre-trained on natural language corpora are a strong option to learn scientific language embeddings through fine-tuning~\cite{christofidellis2023unifying}.

\paragraph{Self Referencing Embedded Strings (SELFIES)}
SELFIES~\cite{krenn2020self} were introduced as an alternative to SMILES to counter the problem of generating invalid molecules. Unlike SMILES, SELFIES are generated using derivation rules to enforce valence-bond validity.
They store branch length and ring size to avoid open branches and rings. These supplementary attributes ensure a valid representation during molecule generation. While this strategy guarantees 100\% validity, it could produce strings that are too short to be a useful molecule. 

\paragraph{International Chemical Identifier (InChI)}
Introduced by the IUPAC, InChI~\cite{heller2015inchi} are strings encoding structural information including charge of the molecule in a hierarchical manner. The strings can get long and complex for larger molecules. To counter this, a hash called ‘InChiKey’ was developed to help with search and retrieval.
InChIs are are less commonly used in LMs~\cite{handsel2021translating}.

\subsection{Generative Modelling}
Generative modeling involves learning the data's underlying distribution with the intent of generating new samples, a technique pivotal in accelerating de novo drug discovery. A generative model may be conditional or unconditional. A conditional generative model utilizes provided data attributes or labels to generate new samples with desired properties, whereas an unconditional model solely provides a way to sample molecules similar to the training data~\cite{gomez2018automatic}. 
The DMTA cycle particularly benefits from the conditional generation approach as it facilitates goal-oriented hypothesis design~\cite{born2021trends}. This section describes a few influential conditional generation models that act on chemical language to generate molecules satisfying user-defined conditions.
\begin{figure}[!htb]
    \centering
    \includegraphics[width=1.0\linewidth]{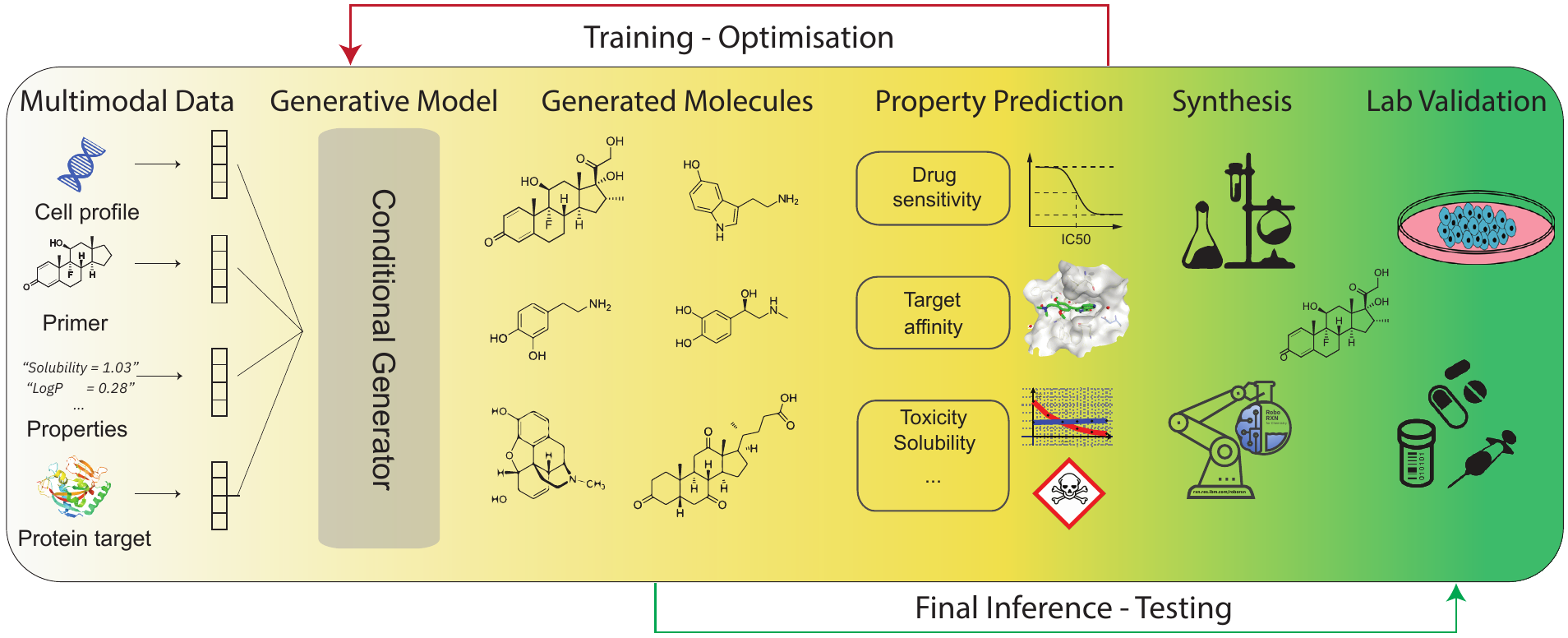}
    \caption{An illustration of conditional molecule generation using LMs. The process initiates with the collection and processing of multi-modal data, which is then compressed into a fixed-size latent representation. These representations are subsequently passed to a molecular generative model. The generated molecules then undergo in-silico property prediction, which is linked back to the generative model through a feedback loop during training. The in-silico models direct the generative model to produce property- or task-driven molecules using a reward function. In the inference stage, candidate molecules generated by the optimized model continue through the workflow for lab synthesis and subsequent experimental validation to determine their efficacy for the desired task.}
    \label{fig:ai_workflow}
\end{figure}


\subsubsection{Recurrent Neural Network (RNN)}
The sequential nature of RNNs makes them suitable models for processing chemical languages. Proposed in the 90s, RNNs were the first flavor of CLMs~\cite{segler2018generating,bjerrum2017smiles,schwaller2018found}.
Their hidden states are continuously updated as new tokens are passed to the network.
During the generation process, tokens are produced auto-regressively. 
 RNNs find use in generating molecule libraries~\cite{segler2018generating} which are extensively used in drug development processes like screening. 
 External scoring functions drive the generation of molecules with desired properties. RNNs are also adept at learning complex distributions~\cite{flam2022language} and generating a higher proportion of unique and valid SMILES~\cite{polykovskiy2020molecular}, even though their inability to count occurrences of ring opening/closing symbols poses a challenge~\cite{joulin2015inferring,popova2018deep}.

\subsubsection{Variational Autoencoder (VAE)}
VAEs learn latent distribution parameters of molecules, thus enabling the generation of new molecules by sampling from this distribution. 
Their unique ability lies in learning a smooth, latent space that facilitates interpolation of samples, even for notoriously discrete entities like molecules~\cite{gomez2018automatic}.
To make it suitable for chemical language models (CLMs), any network compatible with string inputs can function as a VAE's encoder and decoder. Initial works primarily focused on single-modality applications, assessing latent space quality via downstream tasks~\cite{gomez2018automatic}. 
This approach remains prevalent and can be used to generate, e.g., catalysts with an RNN-based VAE~\cite{schilter2023designing} . Here, a latent space is learned and assessed by predicting the catalyst binding energy.
\citet{lim2018molecular} takes it a step further by concatenating a condition vector to the input and the latent embedding generated by the recurrent network-based VAE's encoder. This approach enables the generation of molecules specifically tailored to the given conditions. The scope of VAEs expanded progressively into multi-modal settings for conditional molecule generation, as visualized in~\autoref{fig:ai_workflow} and exemplified by~\citet{born2021paccmannrl, born2021data, born2021active}. These works on task-driven molecule generation incorporate contextual information like gene expression~\cite{born2021paccmannrl} or protein targets~\cite{born2021data,born2021active} or even both~\cite{janakarajan2022fully}. 
VAEs learn embeddings of context information and primer drugs, which are merged before decoding to produce molecules. A reinforcement-learning-based approach directs the model to produce molecules with desired properties using rewards. 

    
\subsubsection{Transformer} The self-attention attribute of Transformers~\cite{vaswani2017attention} have propelled these models to the forefront of NLP. 
Transformers have an encoder module that relies on this self-attention to learn embeddings of the input and the context associated with this input. The decoder module predicts tokens using the context learnt by the encoder and previously generated tokens through attention. For generative modeling, decoder-only transformer like the Generative Pre-Training Transformer (GPT)~\cite{radford2018improving} have become the dominant approach.
This success was translated to the scientific language domain.
One of the first models to use the GPT architecture for conditional molecule generation is MolGPT~\cite{bagal2021molgpt}. SMILES tokens concatenated with a condition vector that summarizes the desired properties and scaffolds are passed as input to this model, which is then trained on the next token prediction task to generate molecules. 
GPT-like models coupled with RL can also be used to optimize molecular properties like pIC50~\cite{mazuz2023molecule}. In this two-stage approach, embeddings are first learnt from SMILES strings, and the embedding space is then optimized such that the model samples molecules with the desired properties. Going beyond just using GPT-like architectures for molecule generation, Regression Transformer~\cite{born2023regression} is a seminal work that formulates conditional sequence modeling as a regression problem.
This gives rise to a natural multitask model that concurrently performs property prediction and conditional molecular generation.
This is achieved by concatenating conventional molecular tokens with property tokens and employing an training scheme that alternates which parts of the sequence are masked.


All these works are testament to the generative capabilities of Transformer-based models. The superior quality of learned embeddings coupled with its ability to handle parallel processing and scalability makes it a top choice for the task of conditional molecule generation, with promising applications in drug discovery and other areas of molecular design~\cite{park2023artificial}.

\subsection{Property Prediction}
Whether a discovery is novel or not, property prediction is a key step in validating the molecules for a given use case. The success of a molecule depends on a myriad of factors, including how it interacts with its environment. 
The MoleculeNet datasets~\cite{wu2018moleculenet} are a commonly used benchmark for property prediction. 
It is curated from public datasets and comprises over 700,000 compounds tested on various properties. 
\citet{born2023chemical} uses a multiscale convolutional attention model to predict toxicity from SMILES. The model has three kernel sizes for the convolutional network and uses a a Bahdanau attention mechanism~\cite{bahdanau2014neural}. The model shows a superior performance overall on various MoleculeNet tasks compared to all other SMILES-based models. 
A recent trend is to use transformer-encoders to learn embeddings for molecules and then apply a multilayer perceptron (MLP) on the embeddings for property prediction. 
MolBERT~\cite{fabian2020molecular} and ChemBERTA~\cite{chithrananda2020chemberta}) are two such examples. These transformer-based models use a BERT backbone to learn molecular embeddings from SMILES and predict properties. 
Similarly, Molformer~\cite{ross2022large} uses a transformer-encoder with linear attention and relative positional encoding to learn compressed molecular representations which are then fine-tuned on chemical property prediction benchmarks. 
To equip transformers with better inductive biases to handle molecules, adaptations of the attention mechanism were proposed.
The molecule attention transformer (MAT) incorporates inter-atomic distances and graph structure into the attention mechanism~\cite{maziarka2019molecule}.
An improvement over this model is the \textit{relative}-MAT  which fuses the distance embedding, bond embedding and neighbourhood embedding and achieves competitive performances on a range of property prediction tasks~\cite{maziarka2021relative}. 

\section{Software tools for scientific language modeling}
The paradigm shift towards open-sourcing software has exerted a profound influence in chemistry.
Commonly listed implications of open-sourcing in the context of drug discovery include catalyzation of methodological development, fostering of collaboration and ease of scientific reproducibility~\cite{gezelter2015open}.
In this section we present several software assets (e.g., Python packages or cloud-based web apps) that are key to enable molecular discovery.

\subsection{Natural language models}
The success story of the Transformer~\cite{vaswani2017attention} as most widely adopted neural network architecture goes hand in hand with the rise of the~\texttt{transformers} library~\cite{wolf2020transformers}, developed since 2019 by~\href{https://huggingface.co}{HuggingFace}. 
Initially intended for NLP applications, Transformers were adopted interdisciplinarily, e.g in computer vision~\cite{dosovitskiy2021image}, reinforcement learning~\cite{chen2021decision}, protein folding~\cite{jumper2021highly} and, of course, chemistry~\cite{schwaller2022machine}.
\textit{HuggingFace} provides the largest public hub of language models and it offers implementations of all recent models as well as a diverse collection of pretrained models available for fine-tuning or inference.
While most of their models focus on NLP, selected models are designed for life science applications, in particular molecular property prediction (e.g.,~\textit{ChemBerta}~\cite{chithrananda2020chemberta}), molecular captioning (e.g.,~\textit{MolT5}~\cite{edwards2022translation}), text-based molecular generation (e.g.,~\textit{MolT5}~\cite{edwards2022translation}) but also unsupervised protein language models (e.g.,~\textit{ProtBert},~\textit{ProtAlbert},~\textit{ProtXLNet} or~\textit{ProtT5}~\cite{elnaggar2021prottrans}).
Moreover, some available models like~\textit{Multimodal Text and Chemistry T5}~\cite{christofidellis2023unifying} are prompt-based multitasker that besides the above mentioned tasks also perform additional tasks such as forward/backward reaction prediction.

\subsection{GT4SD -- Generative modeling toolkits}
Python libraries like~\texttt{GT4SD} (the~\href{https://github.com/GT4SD}{Generative Toolkit for Scientific Discovery}~\cite{manica2023accelerating}), \texttt{TdC} (\href{https://tdcommons.ai}{Therapeutics Data Commons}~\cite{huang2021tdc}) or \texttt{deepchem}~\cite{Ramsundar-et-al-2019} were developed primarily for molecular discovery applications, but especially~\texttt{GT4SD} offers ample support of language models (LMs).
\texttt{GT4SD} is designed to enable researchers and developers to use, train, fine-tune and distribute state-of-the-art generative models for sciences with a focus on the design of organic materials.
It is compatible and inter-operable with many existing libraries and, beyond~\texttt{transformers}, it also gives access to diffusion models (\texttt{diffusers}~\cite{von-platen-etal-2022-diffusers}) or graph generative models (\texttt{TorchDrug}~\cite{zhu2022torchdrug}).
Next to established molecular generation benchmark like~\texttt{Moses}~\cite{polykovskiy2020molecular} and ~\texttt{GuacaMol}~\cite{brown2019guacamol} that include VAEs, generative adversarial networks (GANs), genetic algorithms, and many evaluation metrics for molecular design, \texttt{gt4sd} also supports very contemporary models like the~\textit{Regression Transformer} for concurrent sequence regression and property-driven molecular design~\cite{born2023regression},~\textit{GFlowNets} for highly diverse candidate generation~\cite{bengio2021gflownet} or~\textit{MoLeR} for motif-constrained molecule generation~\cite{maziarz2022learning}.
\texttt{GT4SD} ships with a harmonized interface and a set of command line tools that access a registry of generative models to run or train any model with a few lines of code.
Trained models can be shared to a cloud-hosted model hub and the library is build to facilitate consumption by containerization or distributed computing systems. 
To date, it includes $\sim50$ property prediction endpoints for small molecules, proteins and crystals and overall hosts$~\sim30$ pre-trained algorithms for material design, $20$~\href{https://huggingface.co/GT4SD}{free webapps}~\cite{abid2019gradio} and many Jupyter/Colab notebooks.

\subsection{RXN for Chemistry: Reaction and synthesis language models}
Once a molecule has been selected for experimental validation, a tangible synthesis route has to be identified.
Since the most important tasks in chemical reaction modeling can be framed as sequence conversion problems, the methodology developed for natural language translation can be seamlessly translated to chemistry~\cite{schwaller2022machine}.
In this analogy, atoms are characters, molecules are words, reactions are sentences and precursors are translated into a product or vice versa.

The most mature and flexible library for reaction modeling with LMs is the \href{https://github.com/rxn4chemistry/rxn4chemistry}{package} ~\texttt{rxn4chemistry}~\cite{rxn4chemistry}.
It wraps the API of the \textit{IBM RXN for Chemistry} platform, a \href{https://rxn.res.ibm.com}{freely accessible web application} that gives access to a rich set of language models for different tasks in reaction chemistry.
The flagship architecture has been the \textit{Molecular Transformer} (MT), an autoregressive encoder-decoder model, originally applied to predict outcomes of chemical reactions in organic chemistry~\cite{schwaller2019molecular}.
Notably, the MT uses a purely data-driven, template-free approach that, unlike many graph-based models, can directly represent stereochemistry and thus also exhibits excellent performance on regio- and stereoselective reactions~\cite{pesciullesi2020transfer}. 
The MT was applied to single-step retrosynthesis~\cite{toniato2021unassisted} and became the linchpin of a multi-step retrosynthesis model with a hypergraph exploration strategy~\cite{schwaller2020predicting}.
This approach was later generalized to enzymatic reactions with a tokenization scheme based on enzyme classes which facilitated biocatalyzed synthesis planning and paved the road towards more sustainable and green chemistry~\cite{probst2022biocatalysed}.
Derivatives of the MT helped to enhance diversity in single-step retrosynthesis~\cite{toniato2021unassisted} and a prompt-based disconnection scheme proposed by Thakkar et al.~\cite{thakkar2023unbiasing} significantly improved controllability by allowing the user to mark a disconnection side in the reactant.
Interestingly, an encoder-only derivative of the MT (that replaced the autoregressive decoder with a classification head and leveraged BERT-style~\cite{devlin2018bert} self-supervised pretraining on reactions) excelled in predicting reaction classes~\cite{schwaller2021mapping}.
The hidden representations of such a model were found to encode reaction types and thus allowing to map reaction atlases and to perform reaction similarity search. 
This gave rise to the~\texttt{rxnfp} package for chemical reaction fingerprinting.
Strikingly, masked language modeling also led later to the discovery that the learned attention weights of the Transformer are "secretly" performing atom mapping between products and reactions~\cite{schwaller2021extraction}. 
The epiphany that CLMs accomplish atom mapping without supervision or human labeling bridged the gap between rule-based and data-driven approaches in reaction modeling, making this once tedious experimental task more efficient.

In the quest for automation in organic chemistry, once the precursors for a molecule's synthesis route are identified, the subsequent crucial phase involves seeking an actionable, stepwise synthesis protocol that is ideally amenable for autonomous execution on a robotic platform, such as \textit{IBM RoboRXN}.
In two seminal works Vaucher et al. demonstrated that encoder-decoder Transformers can extract chemical synthesis actions, first from experimental procedures described in patents~\cite{vaucher2020automated} and later predict them directly from the reaction SMILES~\cite{vaucher2021inferring}.
Notable, all the aforementioned models are available via the \textit{IBM RXN for Chemistry} \href{https://rxn.res.ibm.com}{platform} which even allows to control and monitor the robotic platform directly from the web interface. 
For the daunting task of multistep retrosynthesis planning, \textit{RXN} also includes non-transformer based models like \textit{AiZynthFinder}~\cite{genheden2020aizynthfinder}, a Monte Carlo Tree Search approach build on top of a RNN.  
Most of the \textit{RXN} models can be executed also via the \texttt{rxn4chemistry} Python package.

\subsection{Specialized libraries}
\paragraph{Molecular property prediction.}
\texttt{HuggingMolecules} is a library solely devoted to aggregate, standardize and distribute molecular property prediction LMs~\cite{gainski2022huggingmolecules}.
It contains many encoder-only CLMs, some of them with geometrical and structure-aware inductive biases (e.g., the MAT~\cite{maziarka2019molecule} or its successor, the R-MAT~\cite{maziarka2021relative}) while others being pure BERT-based models that were trained on SMILES (e.g,.~\textit{MolBERT}~\cite{fabian2020molecular} or~\textit{ChemBERTA}~\cite{chithrananda2020chemberta}).

\paragraph{Data processing.}
RDKit~\cite{landrum2013rdkit} is a library for manipulating molecules in Python.
For narrower applications like ML data preparation several tools exist.
First,~\texttt{rxn-chemutils} is a \href{https://github.com/rxn4chemistry/rxn-chemutils}{library} with chemistry-related utilities from RXN for Chemistry.
It includes functionalities for standardizing SMILES (e.g., canonicalization or sanitization) but also conversions to other representations (e.g., InChI).
It harmonizes reaction SMILES and prepares them for consumption by CLMs, including also SMILES augmentation (by traversing the molecular graph in a non-canonical order) and tokenization.
Another library with a similar focus is~\texttt{pytoda}~\cite{born2021data,born2021paccmannrl}.
It does not support reaction SMILES but implements richer preprocessing utilities, allowing to chain $>$10 SMILES transformations (e.g., kekulization~\cite{born2023chemical}).
It supports different languages (e.g., SELFIES~\cite{krenn2020self} or BigSMILES~\cite{lin2019bigsmiles}) and tokenization schemes (e.g., SMILES-PE~\cite{li2021smiles}).
Similar functionalities are available for proteins including different languages (IUPAC, UniRep or Blosum62) and protein sequence augmentation strategies~\cite{born2022choice}.
For small molecules, proteins, and polymers, dedicated language classes facilitate the integration with LMs by storing vocabularies, performing online transformations and feeding to custom datasets.
Datasets exist for predicting molecular properties, drug sensitivity, protein-ligand affinity or for self-supervision on small molecules, proteins or polymers.

\subsection{General purpose platforms}
Several general-purpose platforms for molecular discovery have been launched recently, sometimes even preserving privacy through federated learning (i.e., decentralized, distributed training). For example, MELLODDY~\cite{heyndrickx2022melloddy} is a collaborative effort aimed at cross-pharma federated learning of 2.6 billion confidential activity data points.
Similarly, VirtualFlow~\cite{gorgulla2020open} is an open-source platform facilitating large-scale virtual screening that was shown to identify potent KEAP1 inhibitors.
With a focus on \textit{de novo} drug design, Chemistry42~\cite{ivanenkov2023chemistry42} is a proprietary platform integrating AI with computational and medicinal chemistry techniques. 

\section{Future of molecular discovery}
A few years ago, the idea of querying an AI model -- like one would a search engine -- to not only extract scientific knowledge but also perform computational analyses was an overly ambitious feat. Scientific thinking comes from the ability to reason, and AI models cannot reason like humans, yet. However, these models can \textbf{learn} from humans. Our propensity to document everything has enabled us to train Large Language Models (LLMs), like ChatGPT~\cite{chatgpt} and GitHub Copilot~\cite{copilot}, to mimic human responses.
When brought into the context of computational science, this could equip non-experts to confidently conduct computational analyses through well-designed prompts. With human-in-the-loop, a synergistic effect could be created where the scientist provides feedback to the model on its output, thus aiding in better model optimization (a strategy called reinforcement learning from human feedback (RLHF) that has been proven critical for ChatGPT~\cite{christiano2017deep}).
These applications also reduce the barrier for individuals from non-scientific backgrounds to gain a more hands-on experience in conducting scientific analyses without having to go through formal training in computational analysis. 

This section provides a sneak peak into what's next for molecular discovery. Riding the LLM wave, the future holds a place for chatbot-like interfaces that may take care of all things computational in molecular discovery. This includes, for example, generating and iteratively improving design ideas, synthesis planning, material purchasing, performing routine safety checks, and validating experiments.


\subsubsection*{The rise of foundation models in chemistry}
Conventionally, neural networks are trained for a single given task to achieve maximum performance. This essentially renders the models useless for other tasks, thus requiring a new model for every new task, even when the training domain is the same, which in turn imposes a constraint on the rate of our technological advancements. Over the last few years, this conventional approach has been challenged by Large Language Models (LLMs). It has been found that scaling up LLMs leads to astonishing performances in few-shot~\cite{brown2020language} and even zero-shot task generalization~\cite{sanh2022multitask}. Referred to as "foundation models"~\cite{fei2022towards,moor2023foundation}, these models, with typically billions of parameters, can perform multiple tasks despite being trained on one large dataset. Essentially, this multi-task learning is achieved by prompting LLMs with task instructions along with the actual query text which has been found to induce exceptional performance in natural language inference and sentence completion~\cite{sanh2022multitask}. These findings have kicked off new research directions, such as prompt engineering~\cite{wei2022chain} and in-context learning~\cite{brown2020language}, in NLP. 

The foundation model paradigm also finds an increasing adoption in chemistry. There is an increase in task-specific models integrating natural and chemical languages~\cite{vaucher2020automated,vaucher2021inferring,zeng2022deep,edwards2022translation}. 
Concurrently, multi-tasking in pure CLMs has also been advancing through models that combined tasks such as property prediction, reaction prediction and molecule generation either with small task-specific heads (e.g., T5Chem~\cite{lu2022unified}) or via mask infilling (e.g., Regression Transformer~\cite{born2023regression}). \citet{christofidellis2023unifying} were the first to bridge the gap and develop a fully prompt-based multi-task chemical and natural language model.
Despite only 250M parameters, the \textit{Multitask Text and Chemistry T5} was shown to outperform ChatGPT \cite{chatgpt} and Galactica~\cite{taylor2022galactica} on a contrived discovery workflow for re-discovering a common herbicide (natural text $\rightarrow$ new molecule $\rightarrow$ synthesis route $\rightarrow$ synthesis execution protocol).

\subsection{The coalescence of chatbots with chemistry tools}
Given the aforementioned strong task generalization performances of LLMs, building chatbot interfaces around it was a natural next step and thus next to ChatGPT~\cite{chatgpt}, many similar tools were launched.
\begin{figure}[!htb]
    \centering
    \includegraphics[width=1\linewidth]{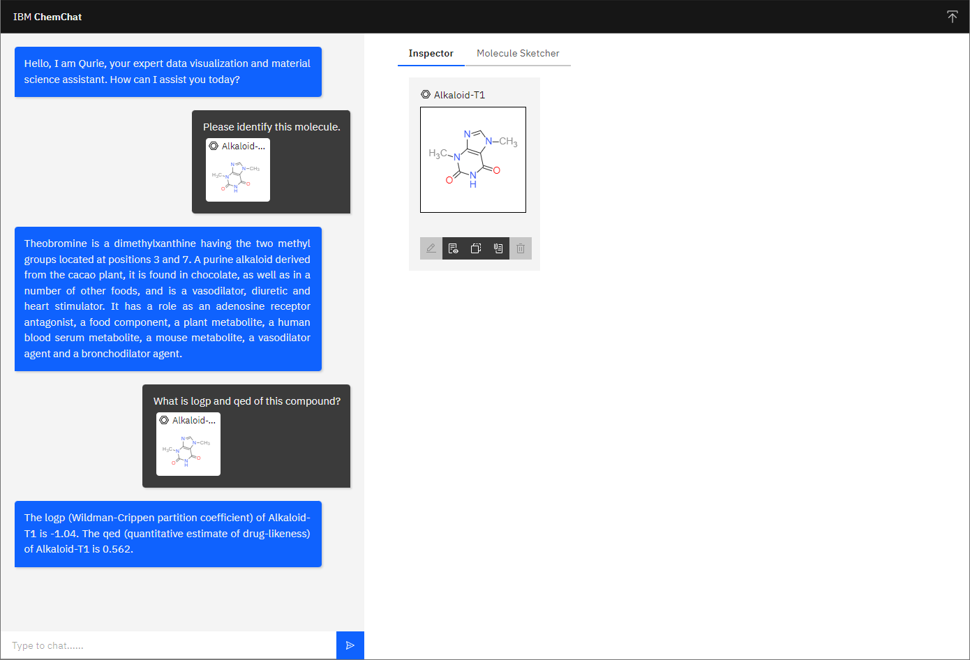}
    \caption{Screenshot of the LLM-powered chatbot application \texttt{ChemChat}. Embedding the capabilities of existing resources such as PubChem~\cite{kim2019pubchem}, RDKit~\cite{landrum2013rdkit} or GT4SD~\cite{manica2023accelerating} enables the assistant to execute programming routines in the background and thus answer highly subject-matter specific user requests without the user needing programming skills.
    }
    \label{fig:ChemChat_01}
\end{figure}
Such tools were found to perform well on simplistic chemistry tasks~\cite{white2023assessment,castro2023large}, opening potential to reshape how chemists interact with chemical data, enabling intuitive access to complex concepts and make valuable suggestions for diverse chemical tasks. 
Furthermore, AI models specifically developed by computer scientists for e.g. drug discovery or material science can be made available through applications powered by LLMs, such as chatbots. 
This minimizes the access barrier for subject matter experts who would otherwise require the respective programming skills to utilize these AI models. 
The power of such chatbots is reached through the coalscence of LLMs and existing chemistry software tools like PubChem~\cite{kim2019pubchem}, RDKit~\cite{landrum2013rdkit} or GT4SD~\cite{manica2023accelerating}. 
Together, such applications can unleash the full potential and value of these models by the strongly enhanced usage. 
An example of how the interaction with such a tool could look like is shown in~\autoref{fig:ChemChat_01}.
In this example, a user provides a molecule (either as SMILES string or via a molecule sketcher) and asks to identify the molecule.
The chatbot relies on prompt-engineering in order to inform the LLM about all its available tools.
The user input is first sent to the LLM which recognizes that one of its supported tools, in this case PubChem, can answer the question.
The chatbot then sends a request to the PubChem API and returns a concise description of the molecule.
The user subsequently asks to compute the logP partition coefficient~\cite{wildman1999prediction} and the quantitative estimate of drug-likeness (QED)~\cite{bickerton2012quantifying}. Calculation of both properties is enabled through the GT4SD tool~\cite{manica2023accelerating} allowing the chatbot to answer the request with certainty.
This will trigger a programming routine to accurately format the API request for GT4SD, i.e., composing the SMILES string with the logP or QED endpoint.
The computation is then performed asynchronously and a separate call to the post-processing routine formats the LLM-generated string reply and composes the response object for the frontend.

\begin{figure}[!htb]
    \centering
    \includegraphics[width=1\linewidth]{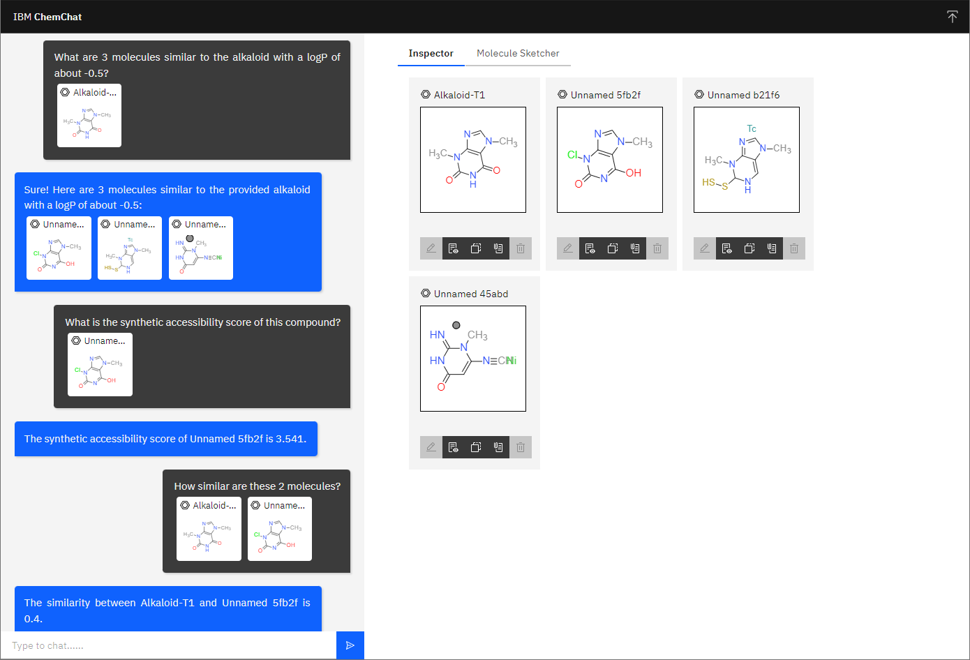}
    \caption{Screenshot of the LLM-powered chatbot application \texttt{ChemChat} showing the continuation of the conversation involving generative tasks through GT4SD's Regression Transformer~\cite{born2023regression} as well as property~\cite{ertl2009estimation} and similarity calculation~\cite{tanimoto1957ibm,rogers2010extended}.
    }
    \label{fig:ChemChat_02}
\end{figure}

This fusion of LLMs with existing tools gives rise to a chatbot assistant for material science and data visualization that can perform simple programming routines without requiring the user to know programming or have access to compute resources.  
A continuation of the conversation involving more complex user queries is shown in~\autoref{fig:ChemChat_02}.
%
%
Having identified the initial molecule as theobromine with a logP of -1.04, the user requests three similar molecules with a slightly increased logP of -0.5.
Here, \texttt{ChemChat} identifies the Regression Transformer~\cite{born2023regression} as the available tool to perform substructure-constrained, property-driven molecule design.
Once the routine has been executed and the three candidate SMILES are collected, the text result is post-processed to add more response data objects such as molecule visualizations, datasets or Vega Lite specs for interactive visualizations.

In conclusion, chatbots can facilitate the integration of essentially all major cheminformatics software in a truly harmonized and seamless manner.
While LLMs are not intrinsically capable to perform complex routines, at least not with high precision and in a trustworthy manner, the synergy between their natural language abilities with existing chemistry tools has the potential to transform the way chemistry is performed.

\FloatBarrier
\bibliographystyle{plainnat}
\bibliography{references}

\begin{thebibliography}{106}
\providecommand{\natexlab}[1]{#1}
\providecommand{\url}[1]{\texttt{#1}}
\expandafter\ifx\csname urlstyle\endcsname\relax
  \providecommand{\doi}[1]{doi: #1}\else
  \providecommand{\doi}{doi: \begingroup \urlstyle{rm}\Url}\fi

\bibitem[cop(2021)]{copilot}
Github copilot.
\newblock \url{https://copilot.github.com/}, 2021.
\newblock Accessed: August 8, 2023.

\bibitem[Abid et~al.(2019)Abid, Abdalla, Abid, Khan, Alfozan, and
  Zou]{abid2019gradio}
Abubakar Abid, Ali Abdalla, Ali Abid, Dawood Khan, Abdulrahman Alfozan, and
  James Zou.
\newblock Gradio: Hassle-free sharing and testing of ml models in the wild.
\newblock \emph{arXiv preprint arXiv:1906.02569}, 2019.

\bibitem[Ar{\'u}s-Pous et~al.(2019)Ar{\'u}s-Pous, Johansson, Prykhodko,
  Bjerrum, Tyrchan, Reymond, Chen, and Engkvist]{arus2019randomized}
Josep Ar{\'u}s-Pous, Simon~Viet Johansson, Oleksii Prykhodko, Esben~Jannik
  Bjerrum, Christian Tyrchan, Jean-Louis Reymond, Hongming Chen, and Ola
  Engkvist.
\newblock Randomized smiles strings improve the quality of molecular generative
  models.
\newblock \emph{Journal of cheminformatics}, 11\penalty0 (1):\penalty0 1--13,
  2019.

\bibitem[Bagal et~al.(2021)Bagal, Aggarwal, Vinod, and
  Priyakumar]{bagal2021molgpt}
Viraj Bagal, Rishal Aggarwal, PK~Vinod, and U~Deva Priyakumar.
\newblock Molgpt: molecular generation using a transformer-decoder model.
\newblock \emph{Journal of Chemical Information and Modeling}, 62\penalty0
  (9):\penalty0 2064--2076, 2021.

\bibitem[Bahdanau et~al.(2014)Bahdanau, Cho, and Bengio]{bahdanau2014neural}
Dzmitry Bahdanau, Kyunghyun Cho, and Yoshua Bengio.
\newblock Neural machine translation by jointly learning to align and
  translate.
\newblock \emph{arXiv preprint arXiv:1409.0473}, 2014.

\bibitem[Bengio et~al.(2021)Bengio, Deleu, Hu, Lahlou, Tiwari, and
  Bengio]{bengio2021gflownet}
Yoshua Bengio, Tristan Deleu, Edward~J Hu, Salem Lahlou, Mo~Tiwari, and
  Emmanuel Bengio.
\newblock Gflownet foundations.
\newblock \emph{Preprint at https://arxiv.org/abs/2111.09266}, 2021.

\bibitem[Bickerton et~al.(2012)Bickerton, Paolini, Besnard, Muresan, and
  Hopkins]{bickerton2012quantifying}
G~Richard Bickerton, Gaia~V Paolini, Jérémy Besnard, Sorel Muresan, and
  Andrew~L Hopkins.
\newblock Quantifying the chemical beauty of drugs.
\newblock \emph{Nat. Chem.}, 4\penalty0 (2):\penalty0 90--98, 2012.

\bibitem[Bjerrum(2017)]{bjerrum2017smiles}
Esben~Jannik Bjerrum.
\newblock Smiles enumeration as data augmentation for neural network modeling
  of molecules.
\newblock \emph{arXiv preprint arXiv:1703.07076}, 2017.

\bibitem[Born and Manica(2021)]{born2021trends}
Jannis Born and Matteo Manica.
\newblock Trends in deep learning for property-driven drug design.
\newblock \emph{Current medicinal chemistry}, 28\penalty0 (38):\penalty0
  7862--7886, 2021.

\bibitem[Born and Manica(2023)]{born2023regression}
Jannis Born and Matteo Manica.
\newblock Regression transformer enables concurrent sequence regression and
  generation for molecular language modelling.
\newblock \emph{Nature Machine Intelligence}, 5\penalty0 (4):\penalty0
  432--444, 2023.

\bibitem[Born et~al.(2021{\natexlab{a}})Born, Huynh, Stroobants, Cornell, and
  Manica]{born2021active}
Jannis Born, Tien Huynh, Astrid Stroobants, Wendy~D Cornell, and Matteo Manica.
\newblock Active site sequence representations of human kinases outperform full
  sequence representations for affinity prediction and inhibitor generation: 3d
  effects in a 1d model.
\newblock \emph{Journal of Chemical Information and Modeling}, 62\penalty0
  (2):\penalty0 240--257, 2021{\natexlab{a}}.

\bibitem[Born et~al.(2021{\natexlab{b}})Born, Manica, Cadow, Markert, Mill,
  Filipavicius, Janakarajan, Cardinale, Laino, and Martínez]{born2021data}
Jannis Born, Matteo Manica, Joris Cadow, Greta Markert, Nil~Adell Mill,
  Modestas Filipavicius, Nikita Janakarajan, Antonio Cardinale, Teodoro Laino,
  and María~Rodríguez Martínez.
\newblock Data-driven molecular design for discovery and synthesis of novel
  ligands: a case study on sars-cov-2.
\newblock \emph{Mach. Learn.: Sci. Technol.}, 2\penalty0 (2):\penalty0 025024,
  2021{\natexlab{b}}.

\bibitem[Born et~al.(2021{\natexlab{c}})Born, Manica, Oskooei, Cadow, Markert,
  and Martínez]{born2021paccmannrl}
Jannis Born, Matteo Manica, Ali Oskooei, Joris Cadow, Greta Markert, and
  María~Rodríguez Martínez.
\newblock Pacc{M}ann\textsuperscript{RL}: De novo generation of hit-like
  anticancer molecules from transcriptomic data via reinforcement learning.
\newblock \emph{iScience}, 24\penalty0 (4):\penalty0 102269,
  2021{\natexlab{c}}.

\bibitem[Born et~al.(2022)Born, Shoshan, Huynh, Cornell, Martin, and
  Manica]{born2022choice}
Jannis Born, Yoel Shoshan, Tien Huynh, Wendy~D Cornell, Eric~J Martin, and
  Matteo Manica.
\newblock On the choice of active site sequences for kinase-ligand affinity
  prediction.
\newblock \emph{Journal of chemical information and modeling}, 62\penalty0
  (18):\penalty0 4295--4299, 2022.

\bibitem[Born et~al.(2023)Born, Markert, Janakarajan, Kimber, Volkamer,
  Mart{\'\i}nez, and Manica]{born2023chemical}
Jannis Born, Greta Markert, Nikita Janakarajan, Talia~B Kimber, Andrea
  Volkamer, Mar{\'\i}a~Rodr{\'\i}guez Mart{\'\i}nez, and Matteo Manica.
\newblock Chemical representation learning for toxicity prediction.
\newblock \emph{Digital Discovery}, 2023.

\bibitem[Brown et~al.(2019)Brown, Fiscato, Segler, and
  Vaucher]{brown2019guacamol}
Nathan Brown, Marco Fiscato, Marwin~HS Segler, and Alain~C Vaucher.
\newblock Guacamol: benchmarking models for de novo molecular design.
\newblock \emph{J. Chem. Inf. Model.}, 59\penalty0 (3):\penalty0 1096--1108,
  2019.

\bibitem[Brown et~al.(2020)Brown, Mann, Ryder, Subbiah, Kaplan, Dhariwal,
  Neelakantan, Shyam, Sastry, Askell, et~al.]{brown2020language}
Tom Brown, Benjamin Mann, Nick Ryder, Melanie Subbiah, Jared~D Kaplan, Prafulla
  Dhariwal, Arvind Neelakantan, Pranav Shyam, Girish Sastry, Amanda Askell,
  et~al.
\newblock Language models are few-shot learners.
\newblock \emph{Advances in neural information processing systems},
  33:\penalty0 1877--1901, 2020.

\bibitem[Castro~Nascimento and Pimentel(2023)]{castro2023large}
Cayque~Monteiro Castro~Nascimento and Andr{\'e}~Silva Pimentel.
\newblock Do large language models understand chemistry? a conversation with
  chatgpt.
\newblock \emph{Journal of Chemical Information and Modeling}, 63\penalty0
  (6):\penalty0 1649--1655, 2023.

\bibitem[Chen et~al.(2021)Chen, Lu, Rajeswaran, Lee, Grover, Laskin, Abbeel,
  Srinivas, and Mordatch]{chen2021decision}
Lili Chen, Kevin Lu, Aravind Rajeswaran, Kimin Lee, Aditya Grover, Misha
  Laskin, Pieter Abbeel, Aravind Srinivas, and Igor Mordatch.
\newblock Decision transformer: Reinforcement learning via sequence modeling.
\newblock \emph{Advances in neural information processing systems},
  34:\penalty0 15084--15097, 2021.

\bibitem[Chithrananda et~al.(2020)Chithrananda, Grand, and
  Ramsundar]{chithrananda2020chemberta}
Seyone Chithrananda, Gabriel Grand, and Bharath Ramsundar.
\newblock Chemberta: large-scale self-supervised pretraining for molecular
  property prediction.
\newblock \emph{arXiv preprint arXiv:2010.09885}, 2020.

\bibitem[Christiano et~al.(2017)Christiano, Leike, Brown, Martic, Legg, and
  Amodei]{christiano2017deep}
Paul~F Christiano, Jan Leike, Tom Brown, Miljan Martic, Shane Legg, and Dario
  Amodei.
\newblock Deep reinforcement learning from human preferences.
\newblock \emph{Advances in neural information processing systems}, 30, 2017.

\bibitem[Christofidellis et~al.(2023)Christofidellis, Giannone, Born, Winther,
  Laino, and Manica]{christofidellis2023unifying}
Dimitrios Christofidellis, Giorgio Giannone, Jannis Born, Ole Winther, Teodoro
  Laino, and Matteo Manica.
\newblock Unifying molecular and textual representations via multi-task
  language modelling.
\newblock In \emph{International Conference on Machine Learning}, 2023.

\bibitem[Das et~al.(2021)Das, Sercu, Wadhawan, Padhi, Gehrmann, Cipcigan,
  Chenthamarakshan, Strobelt, Santos, Chen, et~al.]{das2021accelerated}
Payel Das, Tom Sercu, Kahini Wadhawan, Inkit Padhi, Sebastian Gehrmann, Flaviu
  Cipcigan, Vijil Chenthamarakshan, Hendrik Strobelt, Cicero~Dos Santos, Pin-Yu
  Chen, et~al.
\newblock Accelerated antimicrobial discovery via deep generative models and
  molecular dynamics simulations.
\newblock \emph{Nat. Biomed. Eng.}, 5\penalty0 (6):\penalty0 613--623, 2021.

\bibitem[Devlin et~al.(2018)Devlin, Chang, Lee, and Toutanova]{devlin2018bert}
Jacob Devlin, Ming-Wei Chang, Kenton Lee, and Kristina Toutanova.
\newblock Bert: Pre-training of deep bidirectional transformers for language
  understanding.
\newblock \emph{arXiv preprint arXiv:1810.04805}, 2018.

\bibitem[Dosovitskiy et~al.(2021)Dosovitskiy, Beyer, Kolesnikov, Weissenborn,
  Zhai, Unterthiner, Dehghani, Minderer, Heigold, Gelly, Uszkoreit, and
  Houlsby]{dosovitskiy2021image}
Alexey Dosovitskiy, Lucas Beyer, Alexander Kolesnikov, Dirk Weissenborn,
  Xiaohua Zhai, Thomas Unterthiner, Mostafa Dehghani, Matthias Minderer, Georg
  Heigold, Sylvain Gelly, Jakob Uszkoreit, and Neil Houlsby.
\newblock An image is worth 16x16 words: Transformers for image recognition at
  scale.
\newblock In \emph{9th International Conference on Learning Representations,
  {ICLR} 2021, Virtual Event, Austria, May 3-7, 2021}, 2021.

\bibitem[Edwards et~al.(2022)Edwards, Lai, Ros, Honke, Cho, and
  Ji]{edwards2022translation}
Carl Edwards, Tuan Lai, Kevin Ros, Garrett Honke, Kyunghyun Cho, and Heng Ji.
\newblock Translation between molecules and natural language.
\newblock In \emph{2022 Conference on Empirical Methods in Natural Language
  Processing, EMNLP 2022}, 2022.

\bibitem[Elnaggar et~al.(2021)Elnaggar, Heinzinger, Dallago, Rehawi, Yu, Jones,
  Gibbs, Feher, Angerer, Steinegger, Bhowmik, and Rost]{elnaggar2021prottrans}
Ahmed Elnaggar, Michael Heinzinger, Christian Dallago, Ghalia Rehawi, Wang Yu,
  Llion Jones, Tom Gibbs, Tamas Feher, Christoph Angerer, Martin Steinegger,
  Debsindhu Bhowmik, and Burkhard Rost.
\newblock Prottrans: Towards cracking the language of life's code through
  self-supervised deep learning and high-performance computing.
\newblock \emph{IEEE Transactions on Pattern Analysis and Machine
  Intelligence}, pages 1--1, 2021.
\newblock \doi{10.1109/TPAMI.2021.3095381}.

\bibitem[Ertl and Schuffenhauer(2009)]{ertl2009estimation}
Peter Ertl and Ansgar Schuffenhauer.
\newblock Estimation of synthetic accessibility score of drug-like molecules
  based on molecular complexity and fragment contributions.
\newblock \emph{Journal of cheminformatics}, 1:\penalty0 1--11, 2009.

\bibitem[Fabian et~al.(2020)Fabian, Edlich, Gaspar, Segler, Meyers, Fiscato,
  and Ahmed]{fabian2020molecular}
Benedek Fabian, Thomas Edlich, H{\'e}l{\'e}na Gaspar, Marwin Segler, Joshua
  Meyers, Marco Fiscato, and Mohamed Ahmed.
\newblock Molecular representation learning with language models and
  domain-relevant auxiliary tasks.
\newblock \emph{arXiv preprint arXiv:2011.13230}, 2020.

\bibitem[Fei et~al.(2022)Fei, Lu, Gao, Yang, Huo, Wen, Lu, Song, Gao, Xiang,
  et~al.]{fei2022towards}
Nanyi Fei, Zhiwu Lu, Yizhao Gao, Guoxing Yang, Yuqi Huo, Jingyuan Wen, Haoyu
  Lu, Ruihua Song, Xin Gao, Tao Xiang, et~al.
\newblock Towards artificial general intelligence via a multimodal foundation
  model.
\newblock \emph{Nature Communications}, 13\penalty0 (1):\penalty0 3094, 2022.

\bibitem[Flam-Shepherd et~al.(2022)Flam-Shepherd, Zhu, and
  Aspuru-Guzik]{flam2022language}
Daniel Flam-Shepherd, Kevin Zhu, and Al{\'a}n Aspuru-Guzik.
\newblock Language models can learn complex molecular distributions.
\newblock \emph{Nature Communications}, 13\penalty0 (1):\penalty0 3293, 2022.

\bibitem[for Chemistry~team(2023)]{rxn4chemistry}
IBM~RXN for Chemistry~team.
\newblock {rxn4chemistry: Python wrapper for the IBM RXN for Chemistry API}.
\newblock \url{https://github.com/rxn4chemistry/rxn4chemistry}, 2023.

\bibitem[Gai{\'n}ski et~al.(2022)Gai{\'n}ski, Maziarka, Danel, and
  Jastrzebski]{gainski2022huggingmolecules}
Piotr Gai{\'n}ski, {\L}ukasz Maziarka, Tomasz Danel, and Stanis{\l}aw
  Jastrzebski.
\newblock Huggingmolecules: An open-source library for transformer-based
  molecular property prediction (student abstract).
\newblock In \emph{Proceedings of the AAAI Conference on Artificial
  Intelligence}, volume~36, pages 12949--12950, 2022.

\bibitem[Genheden et~al.(2020)Genheden, Thakkar, Chadimov{\'a}, Reymond,
  Engkvist, and Bjerrum]{genheden2020aizynthfinder}
Samuel Genheden, Amol Thakkar, Veronika Chadimov{\'a}, Jean-Louis Reymond, Ola
  Engkvist, and Esben Bjerrum.
\newblock Aizynthfinder: a fast, robust and flexible open-source software for
  retrosynthetic planning.
\newblock \emph{Journal of cheminformatics}, 12\penalty0 (1):\penalty0 70,
  2020.

\bibitem[Gezelter(2015)]{gezelter2015open}
J~Daniel Gezelter.
\newblock Open source and open data should be standard practices, 2015.

\bibitem[G{\'o}mez-Bombarelli et~al.(2018)G{\'o}mez-Bombarelli, Wei, Duvenaud,
  Hern{\'a}ndez-Lobato, S{\'a}nchez-Lengeling, Sheberla, Aguilera-Iparraguirre,
  Hirzel, Adams, and Aspuru-Guzik]{gomez2018automatic}
Rafael G{\'o}mez-Bombarelli, Jennifer~N Wei, David Duvenaud, Jos{\'e}~Miguel
  Hern{\'a}ndez-Lobato, Benjam{\'\i}n S{\'a}nchez-Lengeling, Dennis Sheberla,
  Jorge Aguilera-Iparraguirre, Timothy~D Hirzel, Ryan~P Adams, and Al{\'a}n
  Aspuru-Guzik.
\newblock Automatic chemical design using a data-driven continuous
  representation of molecules.
\newblock \emph{ACS central science}, 4\penalty0 (2):\penalty0 268--276, 2018.

\bibitem[Gorgulla et~al.(2020)Gorgulla, Boeszoermenyi, Wang, Fischer, Coote,
  Padmanabha~Das, Malets, Radchenko, Moroz, Scott, et~al.]{gorgulla2020open}
Christoph Gorgulla, Andras Boeszoermenyi, Zi-Fu Wang, Patrick~D Fischer, Paul~W
  Coote, Krishna~M Padmanabha~Das, Yehor~S Malets, Dmytro~S Radchenko, Yurii~S
  Moroz, David~A Scott, et~al.
\newblock An open-source drug discovery platform enables ultra-large virtual
  screens.
\newblock \emph{Nature}, 580\penalty0 (7805):\penalty0 663--668, 2020.

\bibitem[Grisoni(2023)]{grisoni2023chemical}
Francesca Grisoni.
\newblock Chemical language models for de novo drug design: Challenges and
  opportunities.
\newblock \emph{Current Opinion in Structural Biology}, 79:\penalty0 102527,
  2023.

\bibitem[Handsel et~al.(2021)Handsel, Matthews, Knight, and
  Coles]{handsel2021translating}
Jennifer Handsel, Brian Matthews, Nicola~J Knight, and Simon~J Coles.
\newblock Translating the inchi: adapting neural machine translation to predict
  iupac names from a chemical identifier.
\newblock \emph{Journal of cheminformatics}, 13\penalty0 (1):\penalty0 1--11,
  2021.

\bibitem[Hargrave-Thomas et~al.(2012)Hargrave-Thomas, Yu, and
  Reynisson]{hargrave2012serendipity}
Emily Hargrave-Thomas, Bo~Yu, and J{\'o}hannes Reynisson.
\newblock Serendipity in anticancer drug discovery.
\newblock \emph{World journal of clinical oncology}, 3\penalty0 (1):\penalty0
  1, 2012.

\bibitem[Heller et~al.(2015)Heller, McNaught, Pletnev, Stein, and
  Tchekhovskoi]{heller2015inchi}
Stephen~R Heller, Alan McNaught, Igor Pletnev, Stephen Stein, and Dmitrii
  Tchekhovskoi.
\newblock Inchi, the iupac international chemical identifier.
\newblock \emph{Journal of cheminformatics}, 7\penalty0 (1):\penalty0 1--34,
  2015.

\bibitem[Heyndrickx et~al.(2022)Heyndrickx, Mervin, Morawietz, Sturm,
  Friedrich, Zalewski, Pentina, Humbeck, Oldenhof, Niwayama,
  et~al.]{heyndrickx2022melloddy}
Wouter Heyndrickx, Lewis Mervin, Tobias Morawietz, No{\'e} Sturm, Lukas
  Friedrich, Adam Zalewski, Anastasia Pentina, Lina Humbeck, Martijn Oldenhof,
  Ritsuya Niwayama, et~al.
\newblock Melloddy: cross pharma federated learning at unprecedented scale
  unlocks benefits in qsar without compromising proprietary information.
\newblock 2022.

\bibitem[Huang et~al.(2021)Huang, Fu, Gao, Zhao, Roohani, Leskovec, W, Xiao,
  Sun, and Zitnik]{huang2021tdc}
Kexin Huang, Tianfan Fu, Wenhao Gao, Yue Zhao, Yusuf Roohani, Jure Leskovec,
  Coley~Connor W, Cao Xiao, Jimeng Sun, and Marinka Zitnik.
\newblock Therapeutics data commons: Machine learning datasets and tasks for
  drug discovery and development.
\newblock \emph{Advances in Neural Information Processing System}, 35, 2021.

\bibitem[Ivanenkov et~al.(2023)Ivanenkov, Polykovskiy, Bezrukov, Zagribelnyy,
  Aladinskiy, Kamya, Aliper, Ren, and Zhavoronkov]{ivanenkov2023chemistry42}
Yan~A Ivanenkov, Daniil Polykovskiy, Dmitry Bezrukov, Bogdan Zagribelnyy,
  Vladimir Aladinskiy, Petrina Kamya, Alex Aliper, Feng Ren, and Alex
  Zhavoronkov.
\newblock Chemistry42: an ai-driven platform for molecular design and
  optimization.
\newblock \emph{Journal of Chemical Information and Modeling}, 63\penalty0
  (3):\penalty0 695--701, 2023.

\bibitem[Janakarajan et~al.(2022)Janakarajan, Born, and
  Manica]{janakarajan2022fully}
Nikita Janakarajan, Jannis Born, and Matteo Manica.
\newblock A fully differentiable set autoencoder.
\newblock In \emph{Proceedings of the 28th ACM SIGKDD Conference on Knowledge
  Discovery and Data Mining}, pages 3061--3071, 2022.

\bibitem[Joulin and Mikolov(2015)]{joulin2015inferring}
Armand Joulin and Tomas Mikolov.
\newblock Inferring algorithmic patterns with stack-augmented recurrent nets.
\newblock \emph{Advances in neural information processing systems}, 28, 2015.

\bibitem[Jumper et~al.(2021)Jumper, Evans, Pritzel, Green, Figurnov,
  Ronneberger, Tunyasuvunakool, Bates, {\v{Z}}{\'\i}dek, Potapenko,
  et~al.]{jumper2021highly}
John Jumper, Richard Evans, Alexander Pritzel, Tim Green, Michael Figurnov,
  Olaf Ronneberger, Kathryn Tunyasuvunakool, Russ Bates, Augustin
  {\v{Z}}{\'\i}dek, Anna Potapenko, et~al.
\newblock Highly accurate protein structure prediction with alphafold.
\newblock \emph{Nature}, 596\penalty0 (7873):\penalty0 583--589, 2021.

\bibitem[Kim et~al.(2019)Kim, Chen, Cheng, Gindulyte, He, He, Li, Shoemaker,
  Thiessen, Yu, et~al.]{kim2019pubchem}
Sunghwan Kim, Jie Chen, Tiejun Cheng, Asta Gindulyte, Jia He, Siqian He,
  Qingliang Li, Benjamin~A Shoemaker, Paul~A Thiessen, Bo~Yu, et~al.
\newblock Pubchem 2019 update: improved access to chemical data.
\newblock \emph{Nucleic acids research}, 47\penalty0 (D1):\penalty0
  D1102--D1109, 2019.

\bibitem[Krenn et~al.(2020)Krenn, H{\"a}se, Nigam, Friederich, and
  Aspuru-Guzik]{krenn2020self}
Mario Krenn, Florian H{\"a}se, AkshatKumar Nigam, Pascal Friederich, and Alan
  Aspuru-Guzik.
\newblock Self-referencing embedded strings (selfies): A 100\% robust molecular
  string representation.
\newblock \emph{Machine Learning: Science and Technology}, 1\penalty0
  (4):\penalty0 045024, 2020.

\bibitem[Landrum(2013)]{landrum2013rdkit}
Greg Landrum.
\newblock Rdkit documentation.
\newblock \emph{Release}, 1\penalty0 (1-79):\penalty0 4, 2013.

\bibitem[Li and Fourches(2020)]{li2020inductive}
Xinhao Li and Denis Fourches.
\newblock Inductive transfer learning for molecular activity prediction:
  Next-gen qsar models with molpmofit.
\newblock \emph{Journal of Cheminformatics}, 12\penalty0 (1):\penalty0 1--15,
  2020.

\bibitem[Li and Fourches(2021)]{li2021smiles}
Xinhao Li and Denis Fourches.
\newblock Smiles pair encoding: a data-driven substructure tokenization
  algorithm for deep learning.
\newblock \emph{Journal of chemical information and modeling}, 61\penalty0
  (4):\penalty0 1560--1569, 2021.

\bibitem[Lim et~al.(2018)Lim, Ryu, Kim, and Kim]{lim2018molecular}
Jaechang Lim, Seongok Ryu, Jin~Woo Kim, and Woo~Youn Kim.
\newblock Molecular generative model based on conditional variational
  autoencoder for de novo molecular design.
\newblock \emph{Journal of cheminformatics}, 10\penalty0 (1):\penalty0 1--9,
  2018.

\bibitem[Lin et~al.(2019)Lin, Coley, Mochigase, Beech, Wang, Wang, Woods,
  Craig, Johnson, Kalow, et~al.]{lin2019bigsmiles}
Tzyy-Shyang Lin, Connor~W Coley, Hidenobu Mochigase, Haley~K Beech, Wencong
  Wang, Zi~Wang, Eliot Woods, Stephen~L Craig, Jeremiah~A Johnson, Julia~A
  Kalow, et~al.
\newblock Bigsmiles: a structurally-based line notation for describing
  macromolecules.
\newblock \emph{ACS central science}, 5\penalty0 (9):\penalty0 1523--1531,
  2019.

\bibitem[Lin et~al.(2023)Lin, Akin, Rao, Hie, Zhu, Lu, Smetanin, Verkuil,
  Kabeli, Shmueli, et~al.]{lin2023evolutionary}
Zeming Lin, Halil Akin, Roshan Rao, Brian Hie, Zhongkai Zhu, Wenting Lu, Nikita
  Smetanin, Robert Verkuil, Ori Kabeli, Yaniv Shmueli, et~al.
\newblock Evolutionary-scale prediction of atomic-level protein structure with
  a language model.
\newblock \emph{Science}, 379\penalty0 (6637):\penalty0 1123--1130, 2023.

\bibitem[Lu and Zhang(2022)]{lu2022unified}
Jieyu Lu and Yingkai Zhang.
\newblock Unified deep learning model for multitask reaction predictions with
  explanation.
\newblock \emph{Journal of Chemical Information and Modeling}, 62\penalty0
  (6):\penalty0 1376--1387, 2022.

\bibitem[Manica et~al.(2023)Manica, Born, Cadow, Christofidellis, Dave, Clarke,
  Teukam, Giannone, Hoffman, Buchan, et~al.]{manica2023accelerating}
Matteo Manica, Jannis Born, Joris Cadow, Dimitrios Christofidellis, Ashish
  Dave, Dean Clarke, Yves Gaetan~Nana Teukam, Giorgio Giannone, Samuel~C
  Hoffman, Matthew Buchan, et~al.
\newblock Accelerating material design with the generative toolkit for
  scientific discovery.
\newblock \emph{npj Computational Materials}, 9\penalty0 (1):\penalty0 69,
  2023.

\bibitem[Maziarka et~al.(2019)Maziarka, Danel, Mucha, Rataj, Tabor, and
  Jastrzkebski]{maziarka2019molecule}
{\L}ukasz Maziarka, Tomasz Danel, S{\l}awomir Mucha, Krzysztof Rataj, Jacek
  Tabor, and S~Jastrzkebski.
\newblock Molecule-augmented attention transformer.
\newblock In \emph{Workshop on Graph Representation Learning, Neural
  Information Processing Systems}, 2019.

\bibitem[Maziarka et~al.(2021)Maziarka, Majchrowski, Danel, Gai{\'n}ski, Tabor,
  Podolak, Morkisz, and Jastrzkebski]{maziarka2021relative}
{\L}ukasz Maziarka, Dawid Majchrowski, Tomasz Danel, Piotr Gai{\'n}ski, Jacek
  Tabor, Igor Podolak, Pawe{\l} Morkisz, and Stanis{\l}aw Jastrzkebski.
\newblock Relative molecule self-attention transformer.
\newblock \emph{arXiv preprint arXiv:2110.05841}, 2021.

\bibitem[Maziarz et~al.(2022)Maziarz, Jackson-Flux, Cameron, Sirockin,
  Schneider, Stiefl, Segler, and Brockschmidt]{maziarz2022learning}
Krzysztof Maziarz, Henry Jackson-Flux, Pashmina Cameron, Finton Sirockin,
  Nadine Schneider, Nikolaus Stiefl, Marwin Segler, and Marc Brockschmidt.
\newblock Learning to extend molecular scaffolds with structural motif.
\newblock In \emph{The Tenth International Conference on Learning
  Representations, {ICLR}}, 2022.

\bibitem[Mazuz et~al.(2023)Mazuz, Shtar, Shapira, and
  Rokach]{mazuz2023molecule}
Eyal Mazuz, Guy Shtar, Bracha Shapira, and Lior Rokach.
\newblock Molecule generation using transformers and policy gradient
  reinforcement learning.
\newblock \emph{Scientific Reports}, 13\penalty0 (1):\penalty0 8799, 2023.

\bibitem[Mikolov et~al.(2013)Mikolov, Chen, Corrado, and
  Dean]{mikolov2013efficient}
Tomas Mikolov, Kai Chen, Greg Corrado, and Jeffrey Dean.
\newblock Efficient estimation of word representations in vector space.
\newblock \emph{arXiv preprint arXiv:1301.3781}, 2013.

\bibitem[Moor et~al.(2023)Moor, Banerjee, Abad, Krumholz, Leskovec, Topol, and
  Rajpurkar]{moor2023foundation}
Michael Moor, Oishi Banerjee, Zahra Shakeri~Hossein Abad, Harlan~M Krumholz,
  Jure Leskovec, Eric~J Topol, and Pranav Rajpurkar.
\newblock Foundation models for generalist medical artificial intelligence.
\newblock \emph{Nature}, 616\penalty0 (7956):\penalty0 259--265, 2023.

\bibitem[OpenAI(2023{\natexlab{a}})]{chatgpt}
OpenAI.
\newblock Chatgpt.
\newblock \url{https://chat.openai.com/chat}, 2023{\natexlab{a}}.
\newblock Accessed: August 8, 2023.

\bibitem[OpenAI(2023{\natexlab{b}})]{openai2023gpt4}
OpenAI.
\newblock Gpt-4 technical report, 2023{\natexlab{b}}.

\bibitem[Park et~al.(2023)Park, Manica, Born, Hedrick, Erdmann, Zubarev,
  Adell-Mill, and Arrechea]{park2023artificial}
Nathaniel~H Park, Matteo Manica, Jannis Born, James~L Hedrick, Tim Erdmann,
  Dmitry~Yu Zubarev, Nil Adell-Mill, and Pedro~L Arrechea.
\newblock Artificial intelligence driven design of catalysts and materials for
  ring opening polymerization using a domain-specific language.
\newblock \emph{Nature Communications}, 14\penalty0 (1):\penalty0 3686, 2023.

\bibitem[Pesciullesi et~al.(2020)Pesciullesi, Schwaller, Laino, and
  Reymond]{pesciullesi2020transfer}
Giorgio Pesciullesi, Philippe Schwaller, Teodoro Laino, and Jean-Louis Reymond.
\newblock Transfer learning enables the molecular transformer to predict
  regio-and stereoselective reactions on carbohydrates.
\newblock \emph{Nature communications}, 11\penalty0 (1):\penalty0 4874, 2020.

\bibitem[Polishchuk et~al.(2013)Polishchuk, Madzhidov, and
  Varnek]{polishchuk2013estimation}
Pavel~G Polishchuk, Timur~I Madzhidov, and Alexandre Varnek.
\newblock Estimation of the size of drug-like chemical space based on gdb-17
  data.
\newblock \emph{J. Comput. Aid. Mol. Des.}, 27\penalty0 (8):\penalty0 675--679,
  2013.

\bibitem[Polykovskiy et~al.(2020)Polykovskiy, Zhebrak, Sanchez-Lengeling,
  Golovanov, Tatanov, Belyaev, Kurbanov, Artamonov, Aladinskiy, Veselov,
  et~al.]{polykovskiy2020molecular}
Daniil Polykovskiy, Alexander Zhebrak, Benjamin Sanchez-Lengeling, Sergey
  Golovanov, Oktai Tatanov, Stanislav Belyaev, Rauf Kurbanov, Aleksey
  Artamonov, Vladimir Aladinskiy, Mark Veselov, et~al.
\newblock Molecular sets (moses): a benchmarking platform for molecular
  generation models.
\newblock \emph{Front. Pharmacol.}, 11:\penalty0 1931, 2020.

\bibitem[Popova et~al.(2018)Popova, Isayev, and Tropsha]{popova2018deep}
Mariya Popova, Olexandr Isayev, and Alexander Tropsha.
\newblock Deep reinforcement learning for de novo drug design.
\newblock \emph{Science advances}, 4\penalty0 (7):\penalty0 eaap7885, 2018.

\bibitem[Probst et~al.(2022)Probst, Manica, Nana~Teukam, Castrogiovanni,
  Paratore, and Laino]{probst2022biocatalysed}
Daniel Probst, Matteo Manica, Yves~Gaetan Nana~Teukam, Alessandro
  Castrogiovanni, Federico Paratore, and Teodoro Laino.
\newblock Biocatalysed synthesis planning using data-driven learning.
\newblock \emph{Nature communications}, 13\penalty0 (1):\penalty0 964, 2022.

\bibitem[Radford et~al.(2018)Radford, Narasimhan, Salimans, Sutskever,
  et~al.]{radford2018improving}
Alec Radford, Karthik Narasimhan, Tim Salimans, Ilya Sutskever, et~al.
\newblock Improving language understanding by generative pre-training.
\newblock 2018.

\bibitem[Ramsundar et~al.(2019)Ramsundar, Eastman, Walters, Pande, Leswing, and
  Wu]{Ramsundar-et-al-2019}
Bharath Ramsundar, Peter Eastman, Patrick Walters, Vijay Pande, Karl Leswing,
  and Zhenqin Wu.
\newblock \emph{Deep Learning for the Life Sciences}.
\newblock O'Reilly Media, 2019.
\newblock
  \url{https://www.amazon.com/Deep-Learning-Life-Sciences-Microscopy/dp/1492039837}.

\bibitem[Rogers and Hahn(2010)]{rogers2010extended}
David Rogers and Mathew Hahn.
\newblock Extended-connectivity fingerprints.
\newblock \emph{Journal of chemical information and modeling}, 50\penalty0
  (5):\penalty0 742--754, 2010.

\bibitem[Ross et~al.(2022)Ross, Belgodere, Chenthamarakshan, Padhi, Mroueh, and
  Das]{ross2022large}
Jerret Ross, Brian Belgodere, Vijil Chenthamarakshan, Inkit Padhi, Youssef
  Mroueh, and Payel Das.
\newblock Large-scale chemical language representations capture molecular
  structure and properties.
\newblock \emph{Nature Machine Intelligence}, 4\penalty0 (12):\penalty0
  1256--1264, 2022.

\bibitem[Sanh et~al.(2022)Sanh, Webson, Raffel, Bach, Sutawika, Alyafeai,
  Chaffin, Stiegler, Le~Scao, Raja, et~al.]{sanh2022multitask}
Victor Sanh, Albert Webson, Colin Raffel, Stephen~H Bach, Lintang Sutawika,
  Zaid Alyafeai, Antoine Chaffin, Arnaud Stiegler, Teven Le~Scao, Arun Raja,
  et~al.
\newblock Multitask prompted training enables zero-shot task generalization.
\newblock In \emph{ICLR 2022-Tenth International Conference on Learning
  Representations}, 2022.

\bibitem[Scannell et~al.(2012)Scannell, Blanckley, Boldon, and
  Warrington]{scannell2012diagnosing}
Jack~W Scannell, Alex Blanckley, Helen Boldon, and Brian Warrington.
\newblock Diagnosing the decline in pharmaceutical r\&d efficiency.
\newblock \emph{Nat. Rev. Drug Discov.}, 11\penalty0 (3):\penalty0 191--200,
  2012.

\bibitem[Schilter et~al.(2023)Schilter, Vaucher, Schwaller, and
  Laino]{schilter2023designing}
Oliver Schilter, Alain Vaucher, Philippe Schwaller, and Teodoro Laino.
\newblock Designing catalysts with deep generative models and computational
  data. a case study for suzuki cross coupling reactions.
\newblock \emph{Digital Discovery}, 2\penalty0 (3):\penalty0 728--735, 2023.

\bibitem[Schwaller et~al.(2018)Schwaller, Gaudin, Lanyi, Bekas, and
  Laino]{schwaller2018found}
Philippe Schwaller, Theophile Gaudin, David Lanyi, Costas Bekas, and Teodoro
  Laino.
\newblock “found in translation”: predicting outcomes of complex organic
  chemistry reactions using neural sequence-to-sequence models.
\newblock \emph{Chemical science}, 9\penalty0 (28):\penalty0 6091--6098, 2018.

\bibitem[Schwaller et~al.(2019)Schwaller, Laino, Gaudin, Bolgar, Hunter, Bekas,
  and Lee]{schwaller2019molecular}
Philippe Schwaller, Teodoro Laino, Th{\'e}ophile Gaudin, Peter Bolgar,
  Christopher~A Hunter, Costas Bekas, and Alpha~A Lee.
\newblock Molecular transformer: a model for uncertainty-calibrated chemical
  reaction prediction.
\newblock \emph{ACS central science}, 5\penalty0 (9):\penalty0 1572--1583,
  2019.

\bibitem[Schwaller et~al.(2020)Schwaller, Petraglia, Zullo, Nair, Haeuselmann,
  Pisoni, Bekas, Iuliano, and Laino]{schwaller2020predicting}
Philippe Schwaller, Riccardo Petraglia, Valerio Zullo, Vishnu~H Nair,
  Rico~Andreas Haeuselmann, Riccardo Pisoni, Costas Bekas, Anna Iuliano, and
  Teodoro Laino.
\newblock Predicting retrosynthetic pathways using transformer-based models and
  a hyper-graph exploration strategy.
\newblock \emph{Chemical science}, 11\penalty0 (12):\penalty0 3316--3325, 2020.

\bibitem[Schwaller et~al.(2021{\natexlab{a}})Schwaller, Hoover, Reymond,
  Strobelt, and Laino]{schwaller2021extraction}
Philippe Schwaller, Benjamin Hoover, Jean-Louis Reymond, Hendrik Strobelt, and
  Teodoro Laino.
\newblock Extraction of organic chemistry grammar from unsupervised learning of
  chemical reactions.
\newblock \emph{Science Advances}, 7\penalty0 (15):\penalty0 eabe4166,
  2021{\natexlab{a}}.

\bibitem[Schwaller et~al.(2021{\natexlab{b}})Schwaller, Probst, Vaucher, Nair,
  Kreutter, Laino, and Reymond]{schwaller2021mapping}
Philippe Schwaller, Daniel Probst, Alain~C Vaucher, Vishnu~H Nair, David
  Kreutter, Teodoro Laino, and Jean-Louis Reymond.
\newblock Mapping the space of chemical reactions using attention-based neural
  networks.
\newblock \emph{Nature machine intelligence}, 3\penalty0 (2):\penalty0
  144--152, 2021{\natexlab{b}}.

\bibitem[Schwaller et~al.(2022)Schwaller, Vaucher, Laplaza, Bunne, Krause,
  Corminboeuf, and Laino]{schwaller2022machine}
Philippe Schwaller, Alain~C Vaucher, Ruben Laplaza, Charlotte Bunne, Andreas
  Krause, Clemence Corminboeuf, and Teodoro Laino.
\newblock Machine intelligence for chemical reaction space.
\newblock \emph{Wiley Interdisciplinary Reviews: Computational Molecular
  Science}, 12\penalty0 (5):\penalty0 e1604, 2022.

\bibitem[Segler et~al.(2018)Segler, Kogej, Tyrchan, and
  Waller]{segler2018generating}
Marwin~HS Segler, Thierry Kogej, Christian Tyrchan, and Mark~P Waller.
\newblock Generating focused molecule libraries for drug discovery with
  recurrent neural networks.
\newblock \emph{ACS central science}, 4\penalty0 (1):\penalty0 120--131, 2018.

\bibitem[Tanimoto(1957)]{tanimoto1957ibm}
Taffee~T Tanimoto.
\newblock Ibm internal report.
\newblock \emph{Nov}, 17:\penalty0 1957, 1957.

\bibitem[Taylor et~al.(2022)Taylor, Kardas, Cucurull, Scialom, Hartshorn,
  Saravia, Poulton, Kerkez, and Stojnic]{taylor2022galactica}
Ross Taylor, Marcin Kardas, Guillem Cucurull, Thomas Scialom, Anthony
  Hartshorn, Elvis Saravia, Andrew Poulton, Viktor Kerkez, and Robert Stojnic.
\newblock Galactica: A large language model for science.
\newblock \emph{arXiv preprint arXiv:2211.09085}, 2022.

\bibitem[Tetko et~al.(2019)Tetko, Karpov, Bruno, Kimber, and
  Godin]{tetko2019augmentation}
Igor~V Tetko, Pavel Karpov, Eric Bruno, Talia~B Kimber, and Guillaume Godin.
\newblock Augmentation is what you need!
\newblock In \emph{International Conference on Artificial Neural Networks},
  pages 831--835. Springer, 2019.

\bibitem[Thakkar et~al.(2023)Thakkar, Vaucher, Byekwaso, Schwaller, Toniato,
  and Laino]{thakkar2023unbiasing}
Amol Thakkar, Alain~C Vaucher, Andrea Byekwaso, Philippe Schwaller, Alessandra
  Toniato, and Teodoro Laino.
\newblock Unbiasing retrosynthesis language models with disconnection prompts.
\newblock \emph{ACS Central Science}, 2023.

\bibitem[Toniato et~al.(2021)Toniato, Schwaller, Cardinale, Geluykens, and
  Laino]{toniato2021unassisted}
Alessandra Toniato, Philippe Schwaller, Antonio Cardinale, Joppe Geluykens, and
  Teodoro Laino.
\newblock Unassisted noise reduction of chemical reaction datasets.
\newblock \emph{Nature Machine Intelligence}, 3\penalty0 (6):\penalty0
  485--494, 2021.

\bibitem[Ucak et~al.(2023)Ucak, Ashyrmamatov, and Lee]{ucak2023improving}
Umit~V Ucak, Islambek Ashyrmamatov, and Juyong Lee.
\newblock Improving the quality of chemical language model outcomes with
  atom-in-smiles tokenization.
\newblock \emph{Journal of Cheminformatics}, 15\penalty0 (1):\penalty0 55,
  2023.

\bibitem[van Deursen et~al.(2020)van Deursen, Ertl, Tetko, and
  Godin]{van2020gen}
Ruud van Deursen, Peter Ertl, Igor~V Tetko, and Guillaume Godin.
\newblock Gen: highly efficient smiles explorer using autodidactic generative
  examination networks.
\newblock \emph{Journal of Cheminformatics}, 12\penalty0 (1):\penalty0 1--14,
  2020.

\bibitem[Vaswani et~al.(2017)Vaswani, Shazeer, Parmar, Uszkoreit, Jones, Gomez,
  Kaiser, and Polosukhin]{vaswani2017attention}
Ashish Vaswani, Noam Shazeer, Niki Parmar, Jakob Uszkoreit, Llion Jones,
  Aidan~N Gomez, {\L}ukasz Kaiser, and Illia Polosukhin.
\newblock Attention is all you need.
\newblock \emph{Advances in neural information processing systems}, 30, 2017.

\bibitem[Vaucher et~al.(2020)Vaucher, Zipoli, Geluykens, Nair, Schwaller, and
  Laino]{vaucher2020automated}
Alain~C Vaucher, Federico Zipoli, Joppe Geluykens, Vishnu~H Nair, Philippe
  Schwaller, and Teodoro Laino.
\newblock Automated extraction of chemical synthesis actions from experimental
  procedures.
\newblock \emph{Nature communications}, 11\penalty0 (1):\penalty0 3601, 2020.

\bibitem[Vaucher et~al.(2021)Vaucher, Schwaller, Geluykens, Nair, Iuliano, and
  Laino]{vaucher2021inferring}
Alain~C Vaucher, Philippe Schwaller, Joppe Geluykens, Vishnu~H Nair, Anna
  Iuliano, and Teodoro Laino.
\newblock Inferring experimental procedures from text-based representations of
  chemical reactions.
\newblock \emph{Nature communications}, 12\penalty0 (1):\penalty0 2573, 2021.

\bibitem[von Platen et~al.(2022)von Platen, Patil, Lozhkov, Cuenca, Lambert,
  Rasul, Davaadorj, and Wolf]{von-platen-etal-2022-diffusers}
Patrick von Platen, Suraj Patil, Anton Lozhkov, Pedro Cuenca, Nathan Lambert,
  Kashif Rasul, Mishig Davaadorj, and Thomas Wolf.
\newblock Diffusers: State-of-the-art diffusion models, 10 2022.
\newblock URL \url{https://github.com/huggingface/diffusers}.

\bibitem[Wei et~al.(2022)Wei, Wang, Schuurmans, Bosma, Xia, Chi, Le, Zhou,
  et~al.]{wei2022chain}
Jason Wei, Xuezhi Wang, Dale Schuurmans, Maarten Bosma, Fei Xia, Ed~Chi, Quoc~V
  Le, Denny Zhou, et~al.
\newblock Chain-of-thought prompting elicits reasoning in large language
  models.
\newblock \emph{Advances in Neural Information Processing Systems},
  35:\penalty0 24824--24837, 2022.

\bibitem[Weininger(1988)]{weininger1988smiles}
David Weininger.
\newblock Smiles, a chemical language and information system. 1. introduction
  to methodology and encoding rules.
\newblock \emph{J. Chem. Inf. Comp. Sci.}, 28\penalty0 (1):\penalty0 31--36,
  1988.

\bibitem[White et~al.(2023)White, Hocky, Gandhi, Ansari, Cox, Wellawatte,
  Sasmal, Yang, Liu, Singh, et~al.]{white2023assessment}
Andrew~D White, Glen~M Hocky, Heta~A Gandhi, Mehrad Ansari, Sam Cox, Geemi~P
  Wellawatte, Subarna Sasmal, Ziyue Yang, Kangxin Liu, Yuvraj Singh, et~al.
\newblock Assessment of chemistry knowledge in large language models that
  generate code.
\newblock \emph{Digital Discovery}, 2\penalty0 (2):\penalty0 368--376, 2023.

\bibitem[Wildman and Crippen(1999)]{wildman1999prediction}
Scott~A Wildman and Gordon~M Crippen.
\newblock Prediction of physicochemical parameters by atomic contributions.
\newblock \emph{Journal of chemical information and computer sciences},
  39\penalty0 (5):\penalty0 868--873, 1999.

\bibitem[Wolf et~al.(2020)Wolf, Debut, Sanh, Chaumond, Delangue, Moi, Cistac,
  Rault, Louf, Funtowicz, et~al.]{wolf2020transformers}
Thomas Wolf, Lysandre Debut, Victor Sanh, Julien Chaumond, Clement Delangue,
  Anthony Moi, Pierric Cistac, Tim Rault, R{\'e}mi Louf, Morgan Funtowicz,
  et~al.
\newblock Transformers: State-of-the-art natural language processing.
\newblock In \emph{Proceedings of the 2020 conference on empirical methods in
  natural language processing: system demonstrations}, pages 38--45, 2020.

\bibitem[Wouters et~al.(2020)Wouters, McKee, and Luyten]{wouters2020estimated}
Olivier~J Wouters, Martin McKee, and Jeroen Luyten.
\newblock Estimated research and development investment needed to bring a new
  medicine to market, 2009-2018.
\newblock \emph{Jama}, 323\penalty0 (9):\penalty0 844--853, 2020.

\bibitem[Wu et~al.(2018)Wu, Ramsundar, Feinberg, Gomes, Geniesse, Pappu,
  Leswing, and Pande]{wu2018moleculenet}
Zhenqin Wu, Bharath Ramsundar, Evan~N Feinberg, Joseph Gomes, Caleb Geniesse,
  Aneesh~S Pappu, Karl Leswing, and Vijay Pande.
\newblock Moleculenet: a benchmark for molecular machine learning.
\newblock \emph{Chemical science}, 9\penalty0 (2):\penalty0 513--530, 2018.

\bibitem[Zeng et~al.(2022)Zeng, Yao, Liu, and Sun]{zeng2022deep}
Zheni Zeng, Yuan Yao, Zhiyuan Liu, and Maosong Sun.
\newblock A deep-learning system bridging molecule structure and biomedical
  text with comprehension comparable to human professionals.
\newblock \emph{Nature communications}, 13\penalty0 (1):\penalty0 862, 2022.

\bibitem[Zhavoronkov et~al.(2019)Zhavoronkov, Ivanenkov, Aliper, Veselov,
  Aladinskiy, Aladinskaya, Terentiev, Polykovskiy, Kuznetsov, Asadulaev,
  et~al.]{zhavoronkov2019deep}
Alex Zhavoronkov, Yan~A Ivanenkov, Alex Aliper, Mark~S Veselov, Vladimir~A
  Aladinskiy, Anastasiya~V Aladinskaya, Victor~A Terentiev, Daniil~A
  Polykovskiy, Maksim~D Kuznetsov, Arip Asadulaev, et~al.
\newblock Deep learning enables rapid identification of potent ddr1 kinase
  inhibitors.
\newblock \emph{Nat. Biotechnol.}, 37\penalty0 (9):\penalty0 1038--1040, 2019.

\bibitem[Zhu et~al.(2022)Zhu, Shi, Zhang, Liu, Xu, Yuan, Zhang, Chen, Cai, Lu,
  et~al.]{zhu2022torchdrug}
Zhaocheng Zhu, Chence Shi, Zuobai Zhang, Shengchao Liu, Minghao Xu, Xinyu Yuan,
  Yangtian Zhang, Junkun Chen, Huiyu Cai, Jiarui Lu, et~al.
\newblock Torchdrug: A powerful and flexible machine learning platform for drug
  discovery.
\newblock \emph{Preprint at https://arxiv.org/abs/2202.08320}, 2022.

\end{thebibliography}
\end{document}